\documentclass[twocolumn,twoside,slac_two]{revtex4}
\usepackage{graphicx}
\usepackage{fancyhdr}
\usepackage{hyperref}
\hypersetup{
    colorlinks=true,
    urlcolor=magenta,
}
\begin{document}
\title{Study of the performance of the HEPD apparatus for the CSES mission}

\author{Beatrice Panico*}
\address{INFN Napoli, I-80126 Naples, Italy - *Corr. author}
\author{Francesco Palma}
\author{Alessandro Sotgiu}
\address{University of Rome Tor Vergata, Via della ricerca scientifica, 00133 Roma, Italy}
\address{INFN Roma Tor Vergata, Via della ricerca scientifica, 00133 Roma, Italy}

\begin{abstract}
The High-Energy Particle Detector (HEPD) is one of the payloads of the CSES space mission. The CSES (China Seismo-Electromagnetic Satellite) mission will investigate the structure and the dynamic of the topside ionosphere, will monitor electric and magnetic field and high energy particle fluctuations, searching for their correlations with the geophysical activity, in order to contribute to the monitoring of earthquakes from space.
The HEPD is built by the Italian collaboration and has different goals. It will study the temporal stability of the inner Van Allen radiation belts, the precipitation of trapped particles in the atmosphere and the low energy component of the cosmic rays (5 - 100 MeV for electrons and 15 - 300 MeV for protons). Here is presented a study of the performance of the apparatus to separate electrons and protons and identify nuclei up to iron.
\end{abstract}

\maketitle

\thispagestyle{fancy}


\section{The CSES Mission}
\label{CSES}
The China Seismo-Electromagnetic Satellite (CSES) is a space mission dedicated to \cite{Wang}:
\begin{enumerate}
 \item measurement of signals from electromagnetic emission and its disturbances in ionosphere;
 \item measurement of background magnetic field in space;
 \item measurement of the disturbance of plasma in ionosphere, such as contents, density and temperature of the ions, density and temperature of the electron;
 \item measurement of energetic particles precipitation.
\end{enumerate}

Some studies \cite{Zhang,Sgrigna,Alex} show a correlation between the electromagnetic anomalies from Earth and ionosphere perturbations and the earthquakes.
In \cite{DeSantis}, a review of the searches for correlation between phenomena in the atmosphere, ionosphere, magnetosphere and seismic events is reported.
Different observables are studied: the precipitation of low energy electrons (E $>$ 0.3 MeV) trapped within the Van Allen Belts \cite{Battiston}; the depletion in the
intensity of VLF electric field at satellite altitudes in \cite{Nemec,Pisa}; time/space localized OLR anomalies \cite{Guo}.

CSES will also complement the cosmic ray measurements in an energy range up to few hundreds of MeV 
and will contribute to the study of the solar-terrestrial environment with the observation of Coronal Mass Ejections. \\

The satellite is based on the Chinese CAST2000 platform. It is a 3-axis attitude stabilized satellite and will be placed in a 97,4$^{\circ}$ 
inclination Sun-synchronous circular orbit, at an altitude of $\sim$500 km. 
The working zone is in the latitude range [-65$^{\circ}$;+65$^{\circ}$], where there is no rotation of solar panel and no action for attitude and orbit
control system of satellite.

The launch is scheduled for 2017, July-August and the expected lifetime is 5 years. \\

CSES hosts several instruments on board, as reported in Tab. \ref{tab:tabCses}.
\begin{table}
 \centering
  \begin{tabular}{ll}
    \hline
    Category & Payloads \\
    \hline
  ElectroMagnetic field & Electric field detector \\
   & High precision magnetometer \\
   & Search coil magnetometer \\
   In-situ plasma & Plasma analyzer package \\
   & Langmuir probe \\
   Plasma construction & GNSS occultation receiver \\
    & Tri-band beacon \\
    Energetic particle & High energy particle package \\
     \hline
  \end{tabular}
  \caption{CSES payloads.}
  \label{tab:tabCses}
\end{table}

\section{The High-Energy Particle Detector}

The High-Energy Particle Detector (HEPD) is one of the payloads of the CSES space mission.
The HEPD is built by the Italian collaboration; its main characteristics are reported in Table \ref{tab:tabHepd}.

\begin{table}
 \centering
  \begin{tabular}{ll}
    \hline
    Parameter & Value\\
    \hline
    Energy Range & Electrons: 3-100 MeV \\
    Energy Range & Protons: 30-200 MeV \\
    Angular resolution & $<8^{\circ}$ at 5 MeV \\
    Energy resolution & $<10\%$ at 5 MeV \\
    Particle identification & $>90\%$ \\
    Free field of view & $\geq 70^{\circ}$ \\
    Pointing & Zenith \\
    Operative temperature & -10$^{\circ}$+45$^{\circ}$ \\
     \hline
  \end{tabular}
  \caption{HEPD main technical characteristics.}
  \label{tab:tabHepd}
\end{table}

To study the low energy component of the cosmic rays, the HEPD detector is composed by different instruments: a tracker, 
made of two planes of double-side silicon micro-strip sensors, at the top of the instrument; the trigger plane made by a plastic scintillator read by 2 PMTs 
and divided into 6 segments; a calorimeter, composed by 16 plastic scintillators with dimensions (15x15x1) cm$^3$ and a layer of 9 LYSO cubes,
for a resulting plane of dimension (15x15x4) cm$^3$; a scintillator veto system, 5 mm thick, at sides and at the bottom of the instrument.

\section{MC Simulation}

\begin{figure}[h!]
\centering
{\includegraphics[width=6cm]{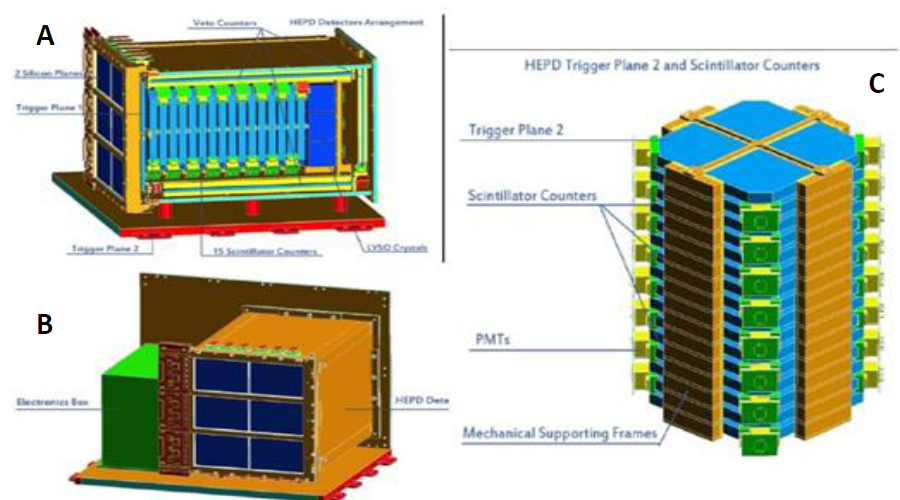}}
\caption[f1]{View of the HEPD electric box and detectors.}
\label{fig:p01}
\end{figure}

A Monte Carlo simulation based on Geant4 toolkit was developed in order to evaluate the HEPD performances. 
In Fig. \ref{fig:p01}A one veto plane has been removed to show the calorimeter inside, the silicon planes are on the left; in Fig. \ref{fig:p01}B the electronic box 
is showed; in Fig. \ref{fig:p01}C the 16 scintillator planes are showed, the first plane is involved in the trigger configuration.  \\

The interaction of particles inside the silicon tracker, the trigger plane, the calorimeter planes and the veto systems, 
togheter with the mechanical structure and the HEPD box were simulated. \\

In order to estimate the performances of HEPD, 10$^9$ events have been uniformly generated according to a power law spectrum $E^{-\gamma}$, where:
\begin{itemize}
 \item $\gamma=2.2$, for electrons in the energy range [1-200] MeV;
 \item $\gamma=2.7$, for protons in the energy range [10-500] MeV;
\end{itemize}

To ensure an uniform exposure for the detector, a spherical surface around HEPD, with a different bin size at the poles and the equator, has been considered.
The simulation has been used to study the trigger mask efficiency, the energy resolution of the calorimeter and the electron/proton discrimination.

\begin{figure}[h!]
\centering
\includegraphics[width=4cm]{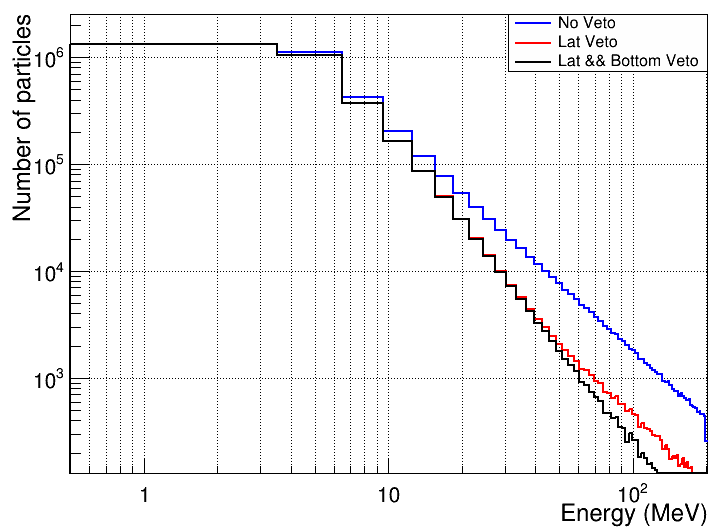}
\includegraphics[width=4cm]{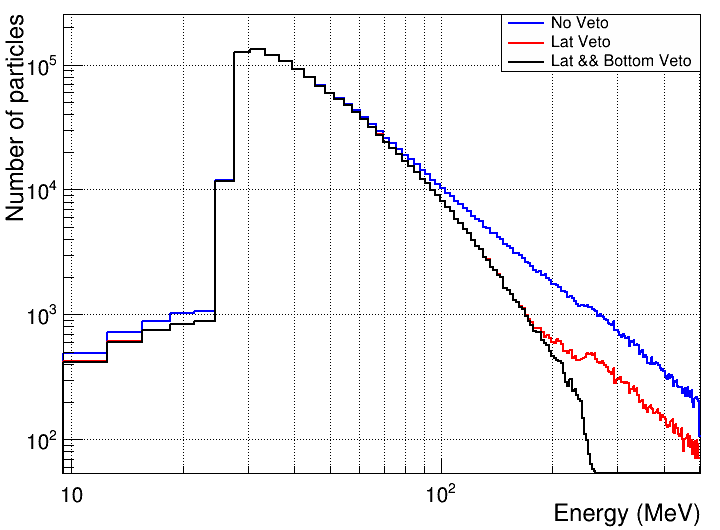}
\caption[f2]{Trigger basic mask (T1 $\&\&$ P1) for simulated electrons (left) and protons (right). The blue line represents all triggered events; the red line represents 
the events without a signal into the lateral veto; the black line represents events completely contained into the detector.}
\label{fig:p02}
\end{figure}

\begin{figure}[h!]
\centering
\includegraphics[width=4cm]{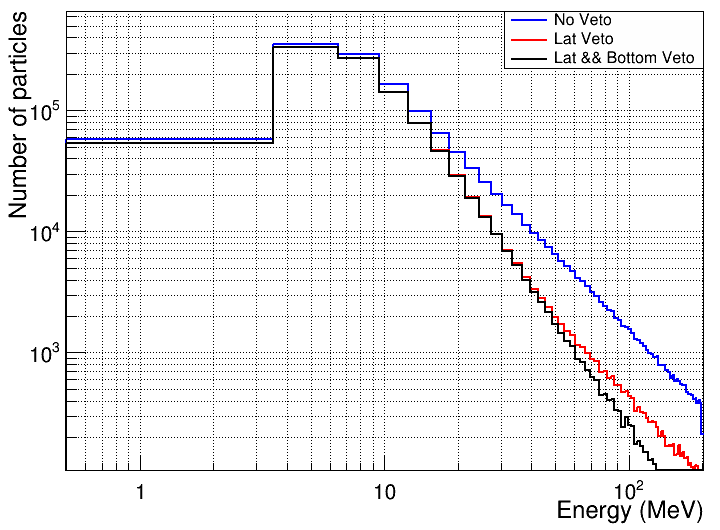}
\includegraphics[width=4cm]{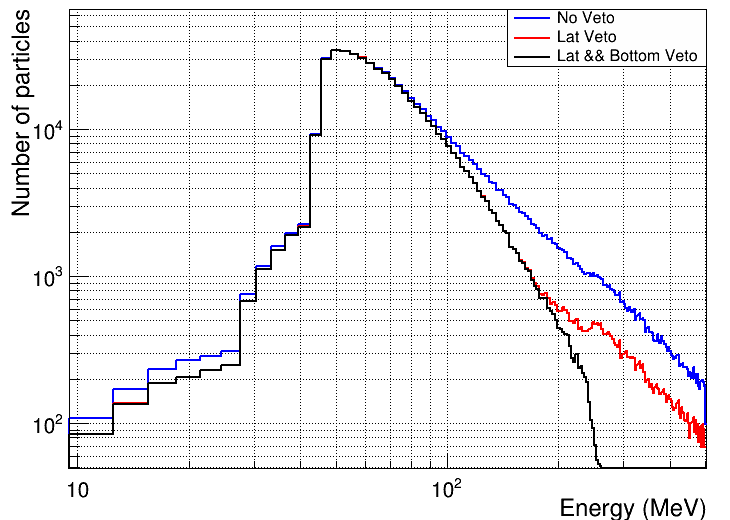}
\caption[f3]{Trigger mask (T1 $\&\&$ P1 $\&\&$ P2) for simulated electrons (left) and protons (right). The blue line represents all triggered events; the red line represents 
the events without a signal into the lateral veto; the black line represents events completely contained into the detector.}
\label{fig:p03}
\end{figure}

In Fig. \ref{fig:p02} the trigger rate is reported with the basic trigger both for electrons and protons. It requires the coincidence between the trigger plane (T1) and the first 
calorimeter plane (P1) (blue line). The red line represents the trigger rate for those events without a signal into the lateral veto; the black line represents the trigger 
rate for events completely contained into the detector.
In Fig. \ref{fig:p03} the trigger rate for a different mask is reported; it requires the coincidence of the trigger plane (T1), of the first (P1) and the second (P2) calorimeter plane.

\subsection{Energy resolution}

\begin{figure}[h!]
\centering
{\includegraphics[width=6cm]{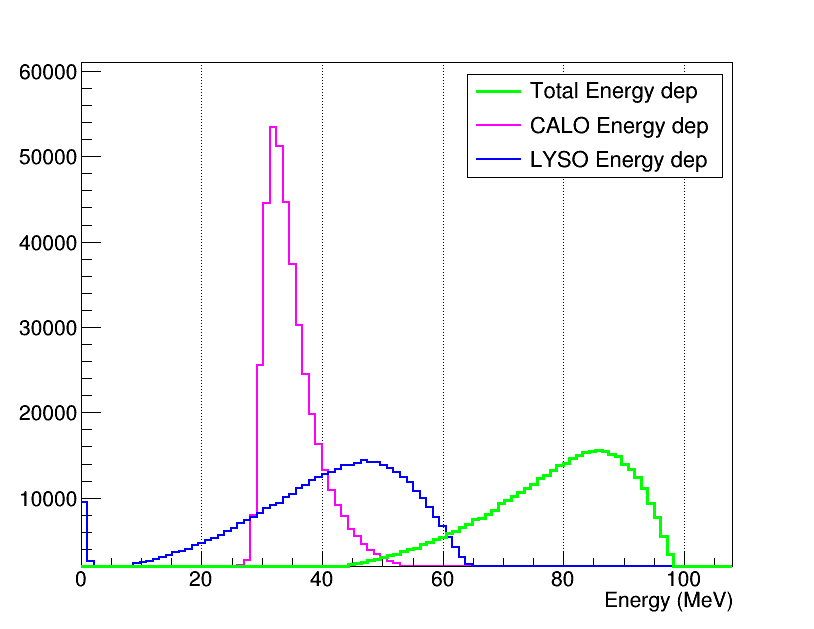}}
{\includegraphics[width=6cm]{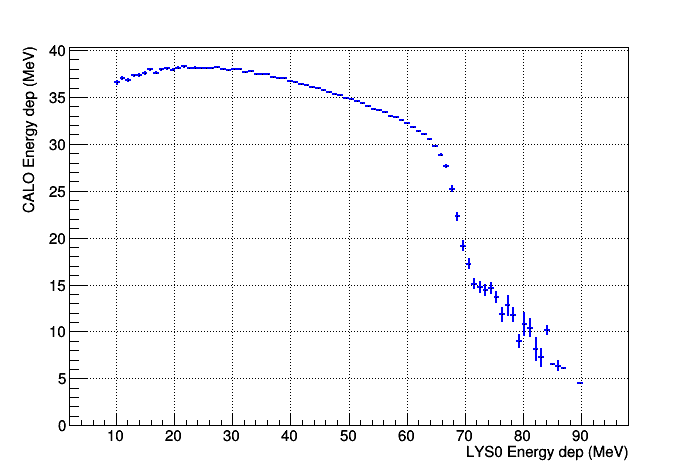}}
\caption[f4]{Energy losses of electrons with energy E=100 MeV into the upper and lower calorimeter (top). 
Profile of the energy released into the upper calorimeter vs the lower calorimeter (down).}
\label{fig:p04}
\end{figure}

\begin{figure}[h!]
\centering
{\includegraphics[width=6cm]{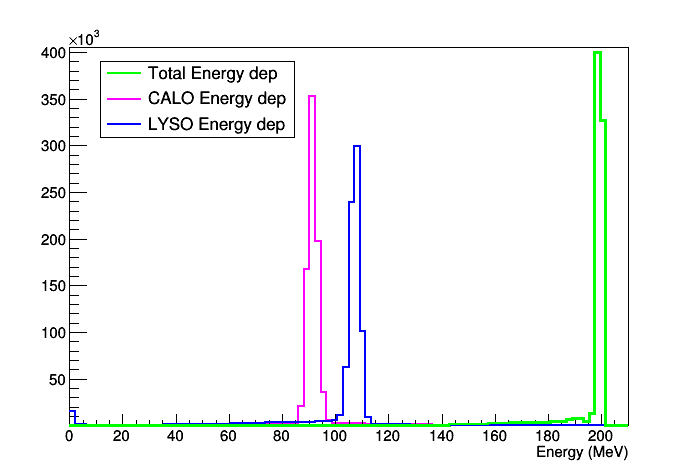}}
{\includegraphics[width=6cm]{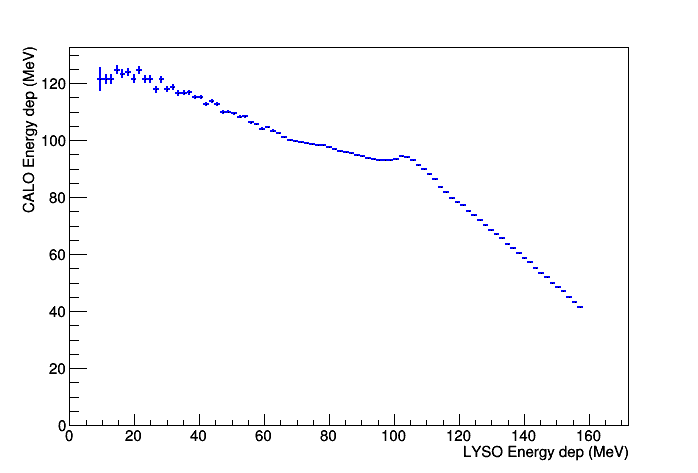}}
\caption[f5]{Energy losses of protons with energy E=200 MeV into the upper and lower calorimeter (top). 
Profile of the energy released into the upper calorimeter vs the lower calorimeter (down).}
\label{fig:p05}
\end{figure}

The HEPD calorimeter is composed by 2 different materials: the upper calorimeter is made by plastic scintillator planes, while the lower calorimeter is made 
by Lyso cristals.
To study the different energy losses in each detector, simulated samples at fixed energies have been used.
The minimum energy required for particles to entirely cross the 3.5 mm aluminium thickness of the satellite wall is $\sim$2 MeV for electrons and $\sim$30 MeV for protons. \\
The simulated energy range is [3-150] MeV for electrons and [30-200] MeV for protons.
In Fig. \ref{fig:p04} the energy losses of electrons with energy E=100 MeV into the upper and lower calorimeter are reported.
The green line represents the total energy released by the particle; the magenta line represents the energy deposited into the upper calorimeter, while the blue line 
is the energy released into the Lyso cristals. In Fig. \ref{fig:p04} the profile of the energy released into the upper calorimeter (CALO) vs the lower calorimeter (LYSO) is also reported.
In Fig. \ref{fig:p05} the same plots are reported for protons with energy E=200 MeV. \\

\begin{figure}[h!]
\centering
{\includegraphics[width=6cm]{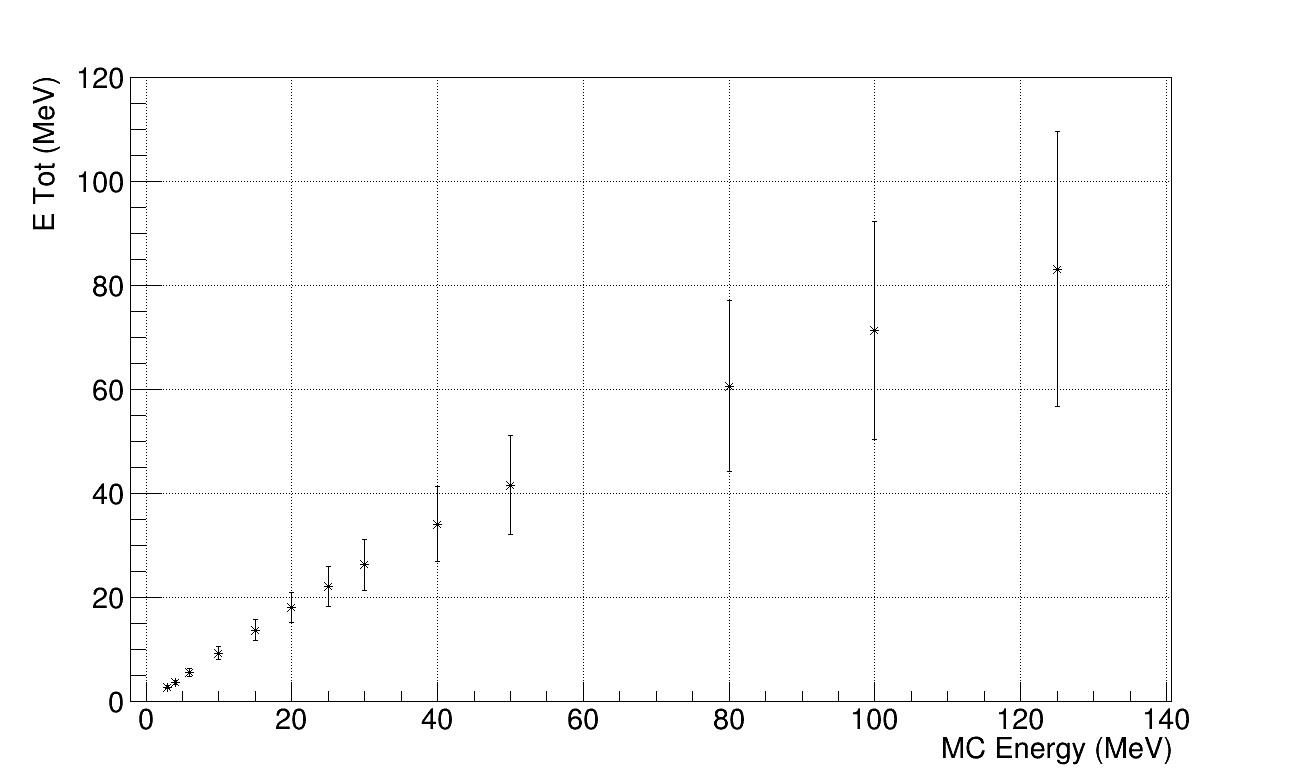}}
{\includegraphics[width=6cm]{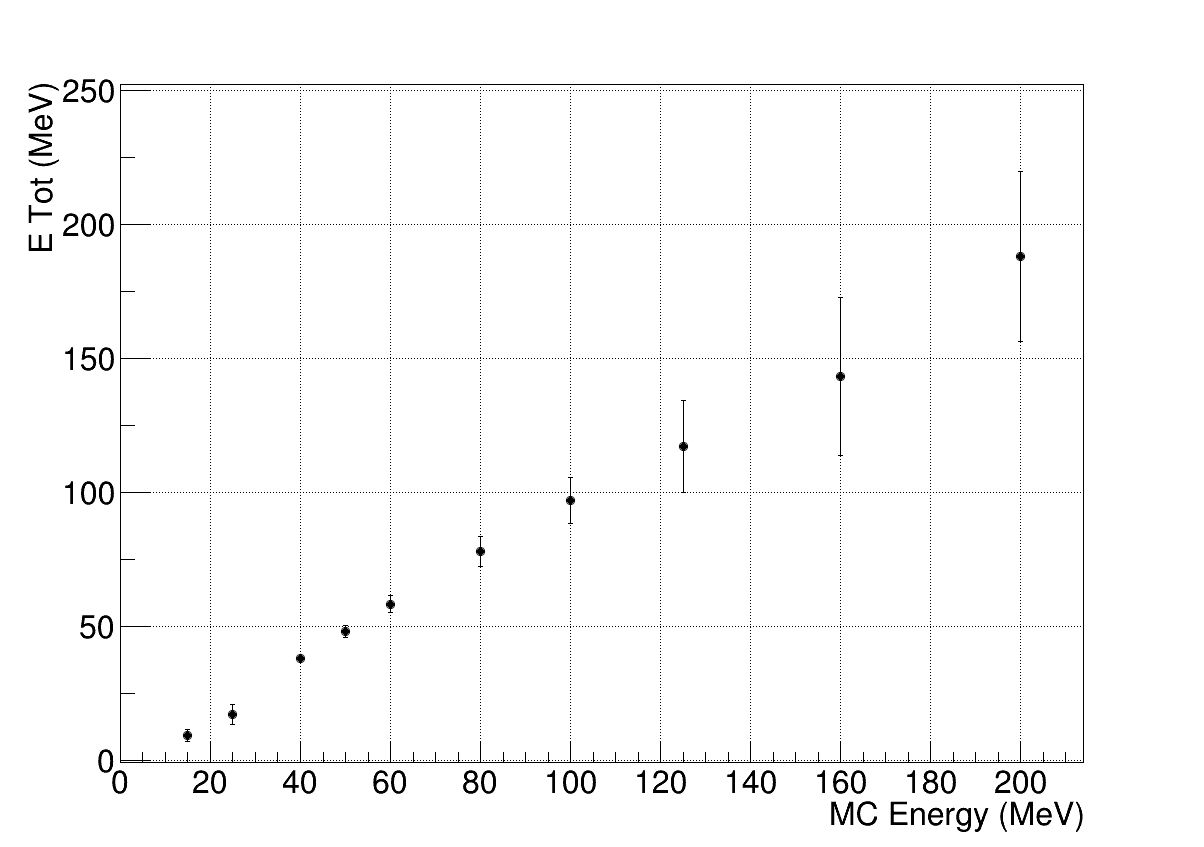}}
\caption[f6]{Energy resolution for the calorimeter for electrons (top) and protons (bottom).}
\label{fig:p06}
\end{figure}

In Fig. \ref{fig:p06} the energy resolution of the HEPD calorimeter is reported. Only events completely contained into the calorimeter have been considered. 
The energy resolution has to be $<10\%$ at 5 MeV. Both for electrons and for protons it is in good agreement with these initial requests.
The energy losses inside the material of the mechanical structures are responsible of the low energy tails observed in the distributions.

\subsection{Electron/Proton discrimination}

One of the most important features of HEPD is the separation between electrons and protons. 

\begin{figure}[h!]
\centering
{\includegraphics[width=6cm]{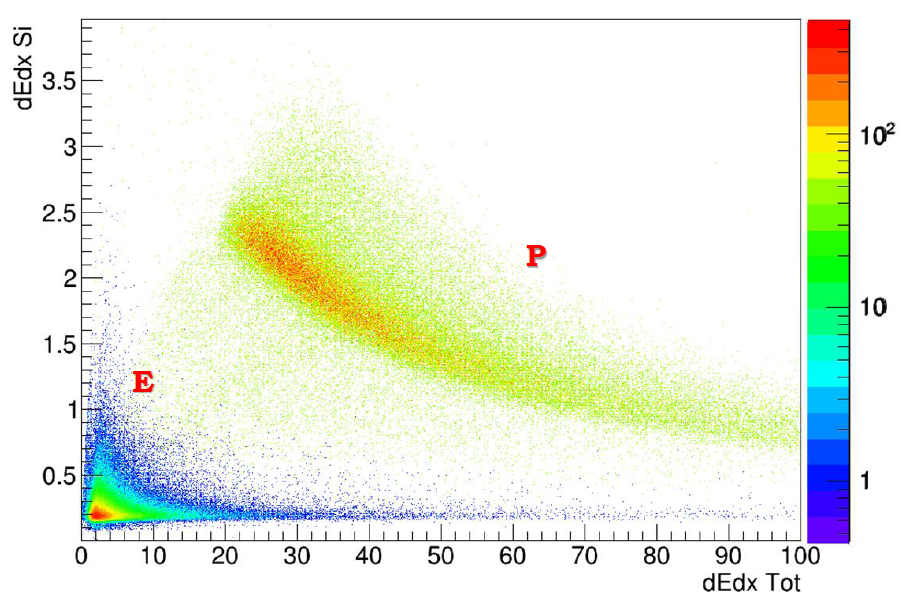}}
{\includegraphics[width=6cm]{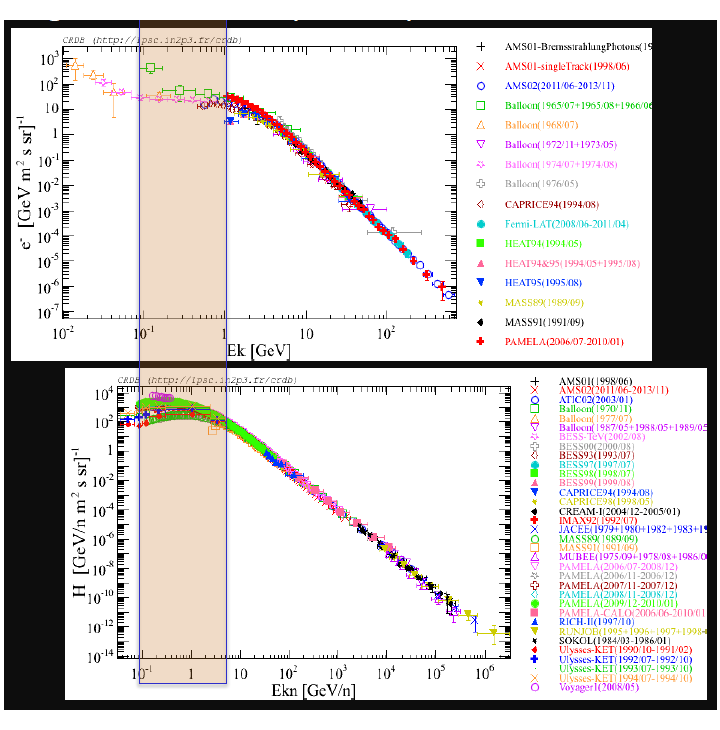}}
\caption[f7]{Top: Distributions of the energy loss in the silicon tracker and the total energy loss in the detector for electrons and protons.
Bottom: Spectrum used to normalize electron and proton MC simulation.}
\label{fig:p07}
\end{figure}

Since, at these energies protons are slow and not relativistic within the HEPD operational range, the \emph{dE~vs~E} method for discriminating 
electrons against protons can be used.
To normalize the simulated proton and electron spectra, the results in \cite{lpsc} have been used.
The normalization factor is chosen at 200 MeV, where proton spectrum is $\sim$45 times greater than the electron spectrum.\\
The \emph{dE} is measured within the two layers of silicon tracker, while the sum of the energy released in the detector provides the \emph{E} measurement.
The veto system allow to select events completely contained into the calorimeter.
The resulting plots for electrons and protons are shown in Fig. \ref{fig:p07}.
Electron and proton distributions are very different and particles lay in separate energy bands.

\section{Conclusions}
In this paper the study of the performances of the HEPD detector have been described. 
The HEPD project involved the construction of 4 detector versions: the Electrical Model, the Structural and Thermal Model, 
the Qualification Model and the Flight Model. \\
The Qualification Model have been tested in April 2016 at the Beam Test Facility (BTF) of the ``Laboratori Nazionali di Frascati" of INFN.
The Flight Model has also been tested at the BTF in Frascati in September 2016 where it was irradiated with electrons and positrons from 30 to 150 MeV. 
In November 2016, the Flight model has been also tested at Trento Facility with a proton beam with energies in the range [35 - 220] MeV.
The beam test data are under analysis. \\

\bigskip 
\nocite{*}


\begin{thebibliography}{9}
\bibitem{Wang} Wang L. et al., Earthq Sci (2015) 28 4, 303
\bibitem{Zhang} Zhang X. et al., Nat. Hazards Earth Syst. Sci. (2013) 13, 197 
\bibitem{Sgrigna} Sgrigna V. et al., Journal of Atmospheric and Solar-Terrestrial Physics (2005), 67 1448
\bibitem{Alex} Alexandrin S. Y. et al., Annales Geophysicae (2003) 21, 597
\bibitem{DeSantis} De Santis A. et al. \emph{Geospace perturbations induced by the Earth: The state of the art and future trends.}, Physics and Chemistry of the Earth, Parts A/B/C, 
(2015) Vol. 85-86, 17 ISSN 1474-7065
\bibitem{Battiston} R. Battiston, V. Vitale, Nuclear Physics B Proceedings Supplement (2013), 243-244, 249
\bibitem{Nemec} Nemec F., Santolik O., Parrot M., Berthelier J. J., Geophys. Res. Lett. (2008), L05109, 35 
\bibitem{Pisa} Pisa D., Nemec F., Parrot M. et al., Annals of Geophysics Volume (2012) 55, Is. 1, 157 
\bibitem{Guo} Guo X. et al., Chinese Journal of GeoPhysics (2010), 53, 6 980
\bibitem{lpsc} http://lpsc.in2p3.fr/crdb/
\end{thebibliography}


\end{document}